\documentclass[11pt]{scrartcl}
\pdfoutput=1

\usepackage{array}
\usepackage{epsfig}
\usepackage{amssymb}
\usepackage{slashed}
\usepackage{amsmath}
\usepackage{graphicx}
\usepackage{cite}
\usepackage{ifpdf}
\usepackage{color}
\usepackage[a4paper, top=1.0in, bottom=1.2in, left=0.8in, right=0.8in]{geometry}

\newcommand{\beq}{\begin{equation}}
\newcommand{\eeq}{\end{equation}}
\newcommand{\beqn}{\begin{eqnarray}}
\newcommand{\eeqn}{\end{eqnarray}}
\newcommand{\be}{\begin{equation}}
\newcommand{\ee}{\end{equation}}
\newcommand{\bea}{\begin{eqnarray}}
\newcommand{\eea}{\end{eqnarray}}
\newcommand{\al}{\alpha}

\newcommand{\om}{\omega}

\newcommand{\cB}{\mathcal B}
\newcommand{\cBSM}{{\mathcal B}^{(0)}_{s, \rm SM}}
\newcommand{\cBSMd}{{\mathcal B}^{(0)}_{d, \rm SM}}

\newcommand{\gev}{{\rm GeV}}
\newcommand{\mev}{{\rm MeV}}

\newcommand{\mm}{\mu \mu}
\newcommand{\BRBsmumu}{{\mathcal B}(B_s \to \mu^+ \mu^-)}
\newcommand{\Bsmumu}{B_s \to \mu^+ \mu^-}
\newcommand{\BRBdmumu}{{\mathcal B}(B_d \to \mu^+ \mu^-)}
\newcommand{\Emax}{E_{\rm max}}
\newcommand{\Bs}{B_s}
\newcommand{\Bd}{B_d}
\newcommand{\MSbar}{\overline{\rm MS}}

\def\sla#1{\setbox0=\hbox{$#1$}\dimen0=\wd0
    \setbox1=\hbox{/} \dimen1=\wd1 \ifdim\dimen0>\dimen1
    \rlap{\hbox to \dimen0{\hfil/\hfil}} #1                        \else
    \rlap{\hbox to \dimen1{\hfil$#1$\hfil}}
    /   \fi}

\begin{document}

\begin{flushright}
{FLAVOUR(267104)-ERC-20}\\
LAPTH-032/12\\
CERN-PH-TH/2012-210
\end{flushright}

\medskip

\begin{center}
{\sffamily \bfseries \LARGE \boldmath On the Standard Model prediction for $\cB(B_{s,d} \to \mu^+ \mu^-)$}\\
[0.8 cm]
{\large Andrzej J. Buras$^a$, Jennifer Girrbach$^a$, Diego~Guadagnoli$^b$, and Gino~Isidori$^{c,d}$}\\
[0.5 cm]
\small
$^a${\em TUM-IAS, Lichtenbergstr. 2a, D-85748 Garching, Germany}\\
[0.1cm]
$^b${\em LAPTh, Universit\'e de Savoie et CNRS, BP110, F-74941 Annecy-le-Vieux Cedex, France}\\
[0.1cm]
$^c${\em CERN, Theory Division, 1211 Geneva 23, Switzerland}\\
[0.1cm]
$^d${\em INFN, Laboratori Nazionali di Frascati, Via E. Fermi 40, 00044 Frascati, Italy}
\end{center}

\vspace{0.3cm}

\abstract{%
\noindent 
The decay $\Bsmumu$ is one of the milestones of the flavor program at the LHC. We reappraise its 
Standard Model prediction. First, by analyzing the theoretical rate in the light of its main 
parametric dependence, we highlight the importance of a complete evaluation of higher-order 
electroweak corrections, at present known only in the large-$m_t$ limit, and leaving sizable 
dependence on the definition of electroweak parameters. Using insights from a complete 
calculation of such corrections for $K\to\pi\nu\bar\nu$ decays, we find a scheme in which 
NLO electroweak corrections are likely to be negligible.
Second, we address the issue of the correspondence between the initial and the final state 
detected by the experiments, and those used in the theoretical prediction. Particular attention 
is devoted to the effect of the soft radiation, that has not been discussed 
for this mode in the previous literature, and that can lead to $O(10\%)$ corrections to the decay 
rate. The ``non-radiative'' branching ratio (that is equivalent to the branching ratio
fully inclusive of bremsstrahlung radiation) is estimated to be 
$(3.23 \pm 0.27) \times 10^{-9}$ for the flavor eigenstate, with the main uncertainty resulting 
from the value of $f_{B_s}$, followed by the uncertainty due to higher order electroweak corrections. 
Applying the same strategy to $B_d\to\mu^+\mu^-$, we find for its non-radiative branching ratio 
$(1.07 \pm 0.10) \times 10^{-10}$.
}

\section{Introduction}\label{sec:intro}

The rare decay $\Bsmumu$ provides one of the best probes of the mechanism of quark-flavor mixing. Within
the Standard Model (SM) this transition is mediated by a flavor-changing neutral current (FCNC) amplitude, 
is helicity suppressed, and is characterized by a purely leptonic final state. The first two features amount 
to a double suppression mechanism, responsible for the extremely rare nature of this decay. The third 
feature causes it to be theoretically very clean at the same time. All these considerations make the 
rare decay $\Bsmumu$ 
a formidable probe of physics beyond the SM, especially of models with a non-standard Higgs sector. The 
$\Bsmumu$ decay has not been observed yet, but the LHC experiments are rapidly approaching the sensitivity 
to observe it~\cite{Aaij:2012ac,Chatrchyan:2012rg,Aad:2012pn} (see also Ref.~\cite{Aaltonen:2011fi}), if it 
occurs at the SM rate. Indeed, the present $95\%$ C.L. bounds read \cite{Aaij:2012ac,Clarke:1429149} 
\be\label{LHCb2}
\mathcal{B}(B_{s}\to\mu^+\mu^-) \le 4.1\times 10^{-9}, \quad
\mathcal{B}(B_{d}\to\mu^+\mu^-) \le 8.2\times 10^{-10}.
\ee
In view of a precise measurement of $\cB(\Bsmumu)$ in the near future, it is of 
utmost importance to assess its SM prediction to the best of our knowledge. 
Analogous comments apply to $B_d\to\mu^+\mu^-$.

In order to obtain a precise prediction for the $\Bsmumu$ rate within the SM it is necessary both to 
compute the corresponding electroweak amplitude with high accuracy and also to assess the correspondence 
between the initial and the final state detected by experiment, and those used in the theoretical 
prediction. More precisely, we can identify three main steps in the comparison between data and theory:
\begin{itemize}

\item{} {\em The evaluation of the ``non-radiative" branching fraction} ($\cB^{(0)}$).
This theoretical quantity is the branching fraction evaluated in the absence of soft-photon corrections.
Thanks to the results of Refs.~\cite{BB-NLO1,BB-NLO2,MU-NLO,BB-NLO3}, $\cB^{(0)}$ is known in the SM including next-to-leading QCD corrections. As a result, it is anticipated that this quantity can be 
computed with an excellent precision, up to the parametric uncertainties from the $B_s$-meson decay 
constant ($f_{B_s}$), the CKM factor ($|V_{tb}^* V_{ts}|$), the $B_s$-meson lifetime ($\tau_{B_s}$), 
and the top-quark mass ($M_t$), in order of decreasing impact on the branching-fraction error. 
Nonetheless, as pointed out in \cite{BB-large-mt}, at the leading order in electroweak corrections 
$\cB^{(0)}$ suffers from a sizable dependence on the renormalization scheme for electroweak 
parameters like $\sin^2\theta_W$ among others. While two-loop electroweak corrections are available
in the large-$m_t$ limit thanks to Ref. \cite{BB-large-mt} itself, we argue that residual uncertainties 
due to the large-$m_t$ approximation are not negligible with respect to the level of accuracy now 
required for the theory prediction. We reassess such corrections, and propose a scheme where they are 
likely to be negligible.

\item{} {\em The treatment of the soft-photon radiation.} 
In the full theory ($\alpha_{\rm em} \not = 0$) photon emission inevitably occurs in this process
(strictly speaking, the width of the non-radiative mode vanishes). The simplest infrared-safe observable 
is 
\beq
\cB^{\rm phys}(\Emax)  \equiv \cB(\Bsmumu + n \gamma)|_{\sum E_\gamma \leq \Emax} 
\eeq
namely the  branching fraction  including an arbitrary number of undetected photons with total energy 
in the meson rest frame less or equal to $\Emax$. As we discuss in section~\ref{sec:IRcorr} (see 
Refs.~\cite{YFS,WeinbergIR,IsidoriIR}), 
\beq
\cB^{\rm phys}(\Emax) = \omega(\Emax) \times \cB^{(0)}~, \label{eq:defomega}
\eeq
where $\omega(\Emax)$ is a correction factor  that we can compute with good accuracy for $\Emax \ll 
m_{B_s}$, and which is independent of possible new-physics contributions affecting $\cB^{(0)}$.
In the limit where we consider bremsstrahlung radiation only,  $\omega(\Emax)$ is known with good 
accuracy for any value of $\Emax$, and in this limit $\omega(\Emax) \to 1$ when $\Emax$ approaches 
its kinematical end point ($\Emax \to m_{B_s}/2$). The theoretical quantity $\cB^{(0)}$ can thus be
identified with the branching ratio {\em fully inclusive} of bremsstrahlung radiation. 

\item{} {\em The time dependence and initial-state tagging.} 
Since the $B_s$ is not a mass eigenstate, also the nature of the initial state and how the measurement 
is performed in time need to be specified. The simplest observable accessible at hadron colliders is 
the flavor-averaged time-integrated distribution. As recently pointed out in 
Refs.~\cite{Fleischer1,Fleischer2}, $\bar B_s - B_s$ oscillation effects do not cancel out in this 
quantity because of the non-vanishing width difference between the two  mass eigenstates.
This leads to a correction factor with respect to the decay rate computed at initial time ($t=0$) 
that, in principle, may be affected by new-physics contributions.
\end{itemize}

The first two points apply also to the $B_d\to\mu^+\mu^-$ decay. On the other hand, in the $B_d$ case 
the complication related to the last point is absent, due to the smallness of $\Delta\Gamma_d$.

In secs. \ref{sec:nonrad} and \ref{sec:IRcorr} we proceed with a detailed discussion of the first 
two points in the case of the $B_s\to\mu^+\mu^-$ decay. Results are then summarized in sec. 
\ref{sec:BR_exp_th}, where they are combined with the third point in order to obtain the SM prediction 
to be compared with the data. The case of $B_d\to\mu^+\mu^-$ is presented in sec. \ref{sec:Bdmumu}.
The final section consists of the list of the main results of our paper and the outlook for the future.

\section{The non-radiative branching ratio} \label{sec:nonrad}

\subsection{Preliminaries}
The SM expression for the branching ratio of the non-radiative decay $B_s\to\mu^+\mu^-$ can be written 
as (see e.g. Ref.~\cite{BBL})
\bea
{\cBSM} &=& \frac{G_F^2   }{\pi} \left[ \frac{\al_{\rm em} (M_Z)}{4\pi \sin^2 \theta_W} \right]^2 
\tau_{\Bs} f_{\Bs}^2 m_{\Bs} m_{\mu}^2 \sqrt{1- \frac{4 m_{\mu}^2}{m_{\Bs}^2}} 
\left| V_{tb}^* V^{\phantom{*}}_{ts} \right|^2 {Y^2( x_{tW}, x_{ht}; \alpha_s)}~,
\label{eq:BR0}
\eea
where $Y$ is an appropriate loop function,\footnote{%
Note that the presence of $\al_{\rm em}$ in the normalization of eq.~(\ref{eq:BR0}) is fictitious: 
we can eliminate it expressing $\al_{\rm em}/\sin^2 \theta_W$ in terms of $G_F$ and $M_W$, thereby 
obtaining an expression that is well defined in the limit $\al_{\rm em}\to 0$.} which 
consists of $Z$-penguin and box-diagram contributions, including QCD corrections as well 
as the leading electroweak corrections. In the absence of such corrections, 
$Y$ reduces to the Inami-Lim function~\cite{InamiLim}
\beq
\label{eq:Y0}
Y_0(x) ~=~ \frac{x}{8}\left( \frac{4-x}{1-x} + \frac{3x}{(1-x)^2} \ln x \right)~,
\eeq
whose argument $x$ can be identified, {in the present discussion,} with
\be
 x_{tW}=\frac{m_t^2(\mu)}{M_W^2}.
\ee
Here $m_t(\mu)$ is the top-quark mass renormalized (as far as 
QCD corrections are concerned) in the $\MSbar$ scheme at the scale $\mu$.

Leaving aside the dominant parametric uncertainty due to $f_{B_s}$ for the time being, 
two evident uncertainties are present in $\cBSM$, if $Y$ is approximated by $Y_0$:
\begin{itemize}

\item
The choice of the scale $\mu$, which is usually chosen to be $O(m_t)$, but
could be as low as $M_W$ or as high as $2m_t$, introducing sizable 
uncertainty in the branching ratio. This unphysical dependence has been 
basically removed through the NLO QCD corrections calculated in 
 Refs.~\cite{BB-NLO1,BB-NLO2,MU-NLO,BB-NLO3}. As these corrections have 
been discussed at length in the literature we will not elaborate on them 
unless necessary. This chapter is closed, at least for this decade.

\item
The choice of renormalization scheme, or equivalently of the definition of 
electroweak parameters, pointed out in ~\cite{BB-large-mt}. 
As this chapter is not yet closed and this uncertainty has not been 
discussed recently in the case of the $B_s\to \mu^+\mu^-$ decay, let us 
have a closer look at this dependence.
\end{itemize}
In order to see that this dependence is sizable, let us compare two definitions 
of $\sin^2\theta_W$, respectively in the $\MSbar$ and in the on-shell scheme,
in which $\sin^2\theta_W$ is very precisely known \cite{PDG}
\be
\label{eq:sW2defs}
\sin^2 \hat\theta_W(M_Z)=0.23116(3), \qquad \left[\sin^2 \theta_W\right]^{\rm OS} 
\equiv 1 - M_W^2/M_Z^2=0.22290~.
\ee
We observe that the second choice, with all other parameters fixed, implies 
$\cBSM$ by $7\%$ higher than the first choice. This corresponds to a shift 
in the rate by $0.22 \times 10^{-9}$, and is equivalent to a shift in $f_{B_s}$ by 
8 MeV, that is larger by almost a factor of two than the error of the most accurate 
determination of this weak decay constant. Evidently this renormalization-scheme
uncertainty has to be removed in the era of precision flavor physics.

The authors of \cite{BB-large-mt} made the first step in this direction 
by providing the result for two-loop electroweak corrections to $Y$ 
in the large-$m_t$ limit. At the level of the branching ratio, this reduced 
the uncertainty by roughly $30\%$, but a warning has been made that 
this estimate could be inaccurate and the inclusion of all NLO electroweak 
corrections will be necessary when the branching ratio in question will be 
precisely measured.
That this estimate could indeed be inaccurate can be seen by keeping only 
the leading term in $m_t$ in $Y_0$, namely $x/8$. This estimate misses the 
true value of $Y_0$ by almost a factor of two.

In what follows we will briefly summarize the findings of \cite{BB-large-mt} 
as far as scheme dependence due to missing NLO electroweak corrections is 
concerned, using the most 
recent set of input parameters. Subsequently we will provide a preliminary solution 
to this problem by using insights from the complete NLO calculation of electroweak 
corrections to $K\to\pi\nu\bar\nu$, that involved the loop function $X_0(x)$. After 
this calculation the remaining uncertainty in $K\to\pi\nu\bar\nu$ related to electroweak 
effects is far below $1\%$ and one should hope that one day this will also be the case 
for $B_s\to\mu^+\mu^-$.

\subsection{Renormalization-scheme dependence}

As already mentioned, we are concerned here with the dependence upon the
choice of the renormalization scheme for electroweak corrections. We consider 
four different renormalization schemes which can be distinguished by the 
manner $\sin^2\theta_W$ and the top-quark mass are renormalized. These are:
\begin{itemize}
\item
Two schemes for $\sin^2\theta_W$ that in the formulae below will be distinguished
by the  parameter $r_s=0,1$:
\be
\sin^2 \hat\theta_W(M_Z):~(r_s=0), 
\qquad \left[\sin^2 \theta_W\right]^{\rm OS}:~(r_s=1),
\ee
with their numerical values given in eq. (\ref{eq:sW2defs}).
\item
Two schemes for the top-quark mass, distinguished by the parameter $r_t=0,1$:
\bea\label{defmt}
m_t \equiv m_t(m_t)^{\rm \MSbar, QCD}:~(r_t=0) \qquad 
\overline m_t \equiv m_t(m_t)^{\rm \MSbar, QCD+EW}:~(r_t=1),
\eea
related via \cite{BB-large-mt}
\be
\overline m_t^2 = m_t^2 \left( 1 + \xi_t \Delta_t(\mu, x_{ht}) \right).
\ee
\end{itemize}
In the case of $m_t$, only QCD corrections are $\MSbar$-renormalized, 
whereas the mass is on-shell as far as electroweak corrections are concerned. 
In the case of $\overline m_t$, both QCD and electroweak corrections are $\MSbar$-renormalized. 
We determine the QCD $\MSbar$ top-quark mass from the pole mass in Table~\ref{tab:input} 
using {\tt RunDec}~\cite{RunDec}.\footnote{~For the central value of $M_t$ in
Table~\ref{tab:input} we obtain $m_t(m_t)^{\rm \MSbar, QCD}= 163.2$ GeV and 
$m_t(m_t)^{\rm \MSbar, QCD+EW}= 164.5$ GeV.}
The explicit expression for $\Delta_t(\mu,x_{ht})$ can be found in Ref.~\cite{BB-large-mt} 
and has been calculated in \cite{Kniehl:1995he}.

We also define
\be\label{defs}
x_{tW}=\frac{m_t^2}{ M_W^2}, \quad
\overline x_{tW}=\frac{{\overline{m}}_t^2}{ M_W^2}, \quad 
x_{ht}=\frac{M_h^2}{m_t^2}, \quad  
\xi_t=\frac{G_F m_t^2}{8 \sqrt2 \, \pi^2}~.
\ee

Concerning other parameters in eq. (\ref{eq:BR0}) we use the Fermi coupling $G_F$ as determined 
from muon decay; $\al_{\rm em} (M_Z)$ denotes the $\MSbar$ QED coupling renormalized at $M_Z$; 
$M_{W,Z}$ are the on-shell masses of the electroweak gauge bosons. 
All the relevant parametric input is collected in Table~\ref{tab:input}.

\long\def\symbolfootnote[#1]#2{\begingroup%
\def\thefootnote{\fnsymbol{footnote}}\footnote[#1]{#2}\endgroup}
\begin{table}[t]
\center{
\begin{tabular}{|l|l|}
\hline
& \\
[-0.35cm]
$G_F = 1.16638 \times 10^{-5}$ GeV$^{-2}$ & $m_{\Bs} = 5.36677$ GeV\\
$\al_{\rm em}^{-1}(M_Z) = 127.937$ \hfill \cite{EWWG_alpha_em} & 
              $f_{\Bs} = 227(8)$ MeV \hfill {\small{[see text]}}\\
$\al_s(M_Z) = 0.1184(7)$ \hfill \cite{Bethke_alpha_s} & $\tau_{\Bs} = 1.466(31)$ ps \\
$M_W = 80.385$ GeV & 
              $|V^*_{tb}V^{\phantom{*}}_{ts}| = 0.0405 (8)\quad $ \hfill \cite{UTfit2012,CKMfitter}\\
$M_Z = 91.1876$ GeV & $m_{\Bd} = 5.27958$ GeV\\
$M_t = 173.2(0.9)$ GeV \hfill \cite{Tevatron_mt_11,Tevatron_mt_12} &
              $f_{\Bd} = 190(8)$ MeV \hfill {\small{[see text]}}\\
$M_h = 125$ GeV \hfill \cite{Higgs} & $\tau_{\Bd} = 1.519(7)$ ps \\
$m_{\mu} = 105.6584$ MeV &
              $|V^*_{tb}V^{\phantom{*}}_{td}| = 0.0087 (2)\quad $ \hfill \cite{UTfit2012,CKMfitter}\\
[0.05cm]
\hline
\end{tabular}
}
\caption{Input parameters used in the determination of $\cBSM$ {and $\cBSMd$}. Quantities without 
an explicit 
reference are taken from Ref.~\cite{PDG}. We do not show the errors for quantities whose uncertainty 
has a negligible impact on our branching-ratio determinations. The central value of 
$f_{B_{s,d}}$ corresponds to the central value of the lattice averages presented in 
Ref.~\cite{Davies:2012qf}, while the error is our estimate of the present uncertainty 
(see text for details).}
\label{tab:input}
\end{table}

Each of the four renormalization schemes in question is characterized by the pair $(r_s,r_t)$. Once this 
pair is fixed, we know uniquely which of the parameters listed above is to be employed in the calculation 
of $Y$ in (\ref{eq:BR0}) and which value of $\sin^2\theta_W$ is to be used in the prefactor in this 
equation. Therefore in presenting a general formula for the function $Y$ valid in all these renormalization 
schemes in the large $m_t$-limit, we can trade the mass variables for the pair $(r_s,r_t)$.

With this notation the loop function $Y$, including complete NLO QCD corrections 
\cite{BB-NLO1,BB-NLO2,MU-NLO,BB-NLO3} and two-loop electroweak corrections in the large-$m_t$ limit 
\cite{BB-large-mt} is given in the $(r_s,r_t)$ scheme as follows:
\beq
\label{eq:Yrsrt}
Y(r_s,r_t; \alpha_s) =  Y_{\rm eff}(r_s,r_t) +  
\frac{\al_s(\mu)}{4\pi} Y_1( x_{tW}),
\eeq
where
\be\label{Yrsrt}
Y_{\rm eff}(r_s,r_t)=Y_0(x_0(r_t)) +  
{\xi_t} \frac{x_{tW}}{8}
\left( \tau_b^{(2)}(x_{ht}) + 3 - 3r_s\frac{\cos^2\theta_W}{\sin^2\theta_W} 
- r_t\Delta_t(\mu,x_{ht})\right)
\ee
with
\be
x_0(r_t) = x_{tW} + r_t (\overline{x}_{tW} - x_{tW})
\ee
is the effective Inami-Lim function for the $(r_s,r_t)$ scheme. This expression 
generalizes the formulae in \cite{BB-large-mt} that applied only to specific 
schemes.
The explicit expression for $\tau_b^{(2)}(x_{ht})$ can be found in  Ref.~\cite{BB-large-mt} 
and has been calculated in \cite{taub2}. Finally, the function 
$Y_1$, encoding the NLO QCD corrections, can be found in \cite{BB-NLO3}. Note 
that $Y_1$ is always evaluated in the $\MSbar$-QCD scheme, that is using 
$x_{tW}$, whereas $Y_0$ is evaluated using $\overline{x}_{tW}$ or $x_{tW}$
depending on the presence or not of the $-\Delta_t$ term in $Y_{\rm eff}$.

In the case of complete NLO electroweak corrections, the $r_s$-dependence in eq. 
(\ref{Yrsrt}) would cancel, up to NNLO effects, the one of $\sin^2\theta_W$ in the prefactor 
in eq. (\ref{eq:BR0}). The corresponding $r_t$ dependence in the correction term in 
(\ref{eq:Yrsrt}) would in turn cancel the one present in the leading term $Y_0$. 
As evident from our formulae, where NLO electroweak corrections are only in the large-$m_t$ 
limit, this cancellation is only partial, implying left-over scheme uncertainties.

Using the central input values in Table~\ref{tab:input} and for 
$\sin^2\theta_W$ in eq. (\ref{eq:sW2defs}) we obtain the central values
for $\cBSM$ in the four renormalization schemes in question, that we 
collect in Table~\ref{tab:schemedep}.  
The central value in either of the cases has been obtained 
setting the QCD renormalization scale to $\mu =  m_t(m_t)^{\rm \MSbar, QCD}.$
We will return to parametric uncertainties at the end of this Section.

The following observations can be made on the basis of this table.
\begin{itemize}
\item
The main remaining uncertainty is due to the choice of the scheme for $\sin^2\theta_W$.
As already found in \cite{BB-large-mt}, the inclusion 
of the NLO electroweak corrections in the large-$m_t$ limit reduced this scheme 
dependence from $7\%$ to $5\%$, but the left-over uncertainty 
is  disturbing.
\item
The left-over uncertainty due to the choice of the scheme for the top-quark 
mass has been reduced to $0.9\%$.
\end{itemize}

\begin{table}[t]
\center{
\begin{tabular}{|l|c|c|}
\hline
& & \\
[-0.35cm]
 &$(r_s,r_t)$& $\cBSM \left[ \times 10^{-9} \right]$ \\
[0.1cm]
\hline
& & \\
[-0.35cm]
$\sin^2 \theta_W$ $\MSbar$,  $m_t$ OS & (0,\,0)  & 3.28 \\
[0.1cm]
$\sin^2 \theta_W$ $\MSbar$, $m_t$ $\MSbar$ & (0,\,1)& 3.31 \\
[0.1cm]
 $\sin^2 \theta_W$ OS, $m_t$ OS & (1,\,0) &  3.42 \\
[0.1cm]
 $\sin^2 \theta_W$ OS, $m_t$ $\MSbar$ &  (1,\,1)&  3.45 \\
[0.05cm]
\hline
\end{tabular}
}
\caption{Dependence of the $\cBSM$ prediction upon the choice of the 
renormalization scheme $(r_s,r_t)$ for electroweak corrections as defined 
in the text.}
\label{tab:schemedep}
\end{table}

In summary the inclusion of the NLO electroweak corrections in the large-$m_t$ limit 
reduced various scheme dependences but the left-over uncertainties are unsatisfactory.
It is also possible that in other schemes the differences could be even larger. Finally, 
one cannot exclude the possibility that, after the inclusion of all NLO electroweak 
corrections, the removal of scheme dependence in the branching ratio would also shift 
significantly its value relatively to the two schemes considered. However, our analysis 
below indicates that for the $(0,0)$ scheme this appears not to be the case.

For completeness, we mention that our results in table~\ref{tab:schemedep} do not 
include log-enhanced QED effects in the RG evolution of the Wilson coefficients
\cite{MM_private}. These corrections, which are part of the full NLO electroweak terms, 
can be calculated by using the results in \cite{ADM_QEDlogs}, and have been included by 
Misiak in his estimate of $\cBSM$ in Ref. \cite{MM_procs}. 
They are found to affect the decay rate by about $-1.4\%$ \cite{MM_private}. Given the 
smallness of these contributions, we prefer not to include them in the absence of a full 
NLO electroweak analysis, comprising also the previously mentioned complete two-loop 
calculation of the electroweak matching conditions.

\subsection{Preliminary solution}

Our analysis shows that, without a complete calculation of NLO electroweak 
effects, only a very rough estimate of the scheme dependence can be made.
At this stage we should emphasize that in all recent papers on $B_s\to \mu^+\mu^-$ 
and most papers in the last decade this uncertainty has been omitted. This can be 
justified by the fact that most authors expected non-SM effects to modify 
the relevant branching ratio by a large amount, rendering any {shift} below $10\%$ 
in the SM estimate irrelevant. With the recent stringent upper bound from LHCb, 
the situation changed dramatically and uncertainties of this size have to be taken 
into account.

Therefore, the question arises, {\em which value for $\cBSM$ should be quoted in 
the absence of complete NLO electroweak corrections.}

Here we would like to propose  a preliminary solution to this problem. 
As already pointed out in \cite{BB-large-mt} the same problem is present in 
$K\to\pi\nu\bar\nu$ decays, which are theoretically even cleaner than 
$B_s\to\mu^+\mu^-$. These decays are governed by the Inami-Lim function 
$X_0(x_t)$, which differs from $Y_0(x_t)$ only by box contributions. Therefore 
at large $m_t$, where only the $Z$-penguin is relevant, the effective electroweak 
corrections to Inami-Lim functions are identical to the ones presented 
above.

Now comes an important point. We are {in the lucky circumstance} that complete 
NLO electroweak corrections to $K\to\pi\nu\bar\nu$ have been calculated by 
Brod, Gorbahn and Stamou two years ago \cite{Brod:2010hi}. These authors considered 
three renormalization schemes:
\begin{itemize}
\item
The $\MSbar$ scheme for all parameters. In our terminology this is the (0,1) scheme.
\item
The $\MSbar$ scheme for all couplings and the on-shell scheme for all masses. 
This is the (0,0) scheme.
\item
The on-shell scheme for weak mixing angle and all masses and the QED coupling 
constant renormalized in the  $\MSbar$ scheme. This is the (1,0) scheme.
\end{itemize}
By calculating complete NLO electroweak corrections in these three schemes,
they reduced the scheme dependence at the level of the branching ratio 
far below $1\%$, a remarkable result. Looking at the size of different corrections 
they concluded that the on-shell definition of masses, together with the $\MSbar$ 
definition of $\sin^2\theta_W$, our (0,0) scheme, is the best choice of the 
renormalization scheme, namely the scheme where NLO corrections
are smallest in absolute value. Incidentally, we find that, in our $\Bsmumu$ case, 
this scheme is also the one that exhibits the smallest dependence, below 1\%,
upon the choice of the renormalization scale in the range $[M_Z,m_t]$.

By inspection of their analysis for the mentioned scheme, in particular of equations 
(4.2) -- (4.4) of their paper, a very simple prescription for the final result 
for $K\to\pi\nu\bar\nu$ branching ratios (including complete NLO QCD and complete 
NLO electroweak corrections) {emerges}. Adapted to the $B_s\to\mu^+\mu^-$ decay, 
this prescription is as follows:
\begin{itemize}
\item 
 Use eq. (\ref{eq:BR0}) for $\cBSM$ with 
\be\label{CONJ1}
\sin^2 \theta_W =\sin^2 \hat\theta_W(M_Z)=0.23116(3).
\ee
\item
Set
\be
\label{eq:Y0prescription}
Y(x_{tW}, x_{ht}; \alpha_s)=Y_0(x_{tW}) + 
\frac{\al_s(\mu)}{4\pi} Y_1( x_{tW})\equiv \eta_Y Y_0(x_{tW}),\qquad \eta_Y = 1.0113~,
\ee
\end{itemize}
where $x_{tW}$ is defined by eqs. (\ref{defmt}) and (\ref{defs}).
Our value of $\eta_Y$ agrees well with 1.012 quoted in \cite{BB-NLO3}.

The complete electroweak corrections to $B_s\to\mu^+\mu^-$ will be 
different in the details, due to different box diagrams and the presence of
charged leptons in the final state in place of neutrinos. Yet it is 
plausible to expect that the prescription given above could work here 
as well.

We emphasize that the prescription described here can only be validated
by a full-fledged NLO calculation of electroweak corrections. Indeed, in the case of
$K\to\pi\nu\bar\nu$ the adherence of our simple prescription to the full
NLO result can be checked because of the existence of such complete calculation, 
thanks to Ref. \cite{Brod:2010hi}. It is known from any perturbative calculation, both 
in QCD and electroweak theory, that a particular definition of fundamental
parameters and renormalization scale in the leading term allows to 
minimize NLO corrections. Here we are conjecturing, based on the 
argument presented at the beginning of this section, that the same choice 
that works for $K\to\pi\nu\bar\nu$ will plausibly work for $B_s\to\mu^+\mu^-$ 
as well. A complete analysis of NLO electroweak corrections for the latter 
decay is the only way to confirm {our conjecture}.

\subsection{Final result}

Using the prescription given above we now calculate $\cBSM$, with 
the plausible expectation (in the sense discussed above) that scheme dependence 
is kept to a minimum.
As anticipated, in this case the dominant source of uncertainty in $\cBSM$ 
is of parametric nature. Treating all errors as Gaussian, we obtain the final result
\be
\label{eq:B0num_ref}
{
 \cBSM =(3.23 \pm 0.27)  \times 10^{-9},
 }
\ee
which is closest to our large-$m_t$ result in the (0,0) scheme.
We choose eq. (\ref{eq:B0num_ref}) as our reference value for the non-radiative
branching ratio.

For future reference, we illustrate the impact of the various inputs via the following 
parametric expression:
\bea
\label{eq:BR0par}
&& \cBSM =  {3.2348} \times 10^{-9} \times \left( \frac{M_t}{173.2~\gev} \right)^{3.07}
\left( \frac{f_{B_s}}{227~\mev} \right)^2
\left( \frac{\tau_{B_s}}{1.466~\rm{ps}}\right)
\left| \frac{V^*_{tb} V^{\phantom{*}}_{ts} }{4.05 \times 10^{-2}} \right|^2 \nonumber  \\
&& \qquad =   \left( 
{3.23}\pm  0.15 \pm 0.23_{f_{B_s}} 
\right)   \times 10^{-9}~. \eea
In the second line of eq.~(\ref{eq:BR0par}) we have explicitly separated the contribution to the error 
due to $f_{B_s}$, which is the most relevant source of uncertainty and deserves a dedicated discussion. 

As pointed out in Ref.~\cite{Buras:2003td}, in principle one can get rid of the quadratic $f_{B_s}$ 
dependence in $\cB(B_s\to \mu^+\mu^-)$ by normalizing this observable to $\Delta m_{B_s}$, thereby 
taking advantage of the relatively precise lattice results on the bag parameter of the 
$\bar B_s - B_s$ mixing amplitude, that enters the latter linearly. Moreover, this procedure 
removes also the dependence on the CKM parameters. Indeed, in 2003 this proposal reduced the 
uncertainty in $\cB(B_s\to \mu^+\mu^-)$ by a factor of three.
However, given the recent progress in the direct determination of $f_{B_s}$ from the lattice 
\cite{Davies:2012qf,Dimopoulos:2011gx,McNeile:2011ng,Bazavov:2011aa,Blossier:2011dk,Na:2012kp,%
Gamiz:2009ku,Bouchard:2011xj} and in the determination of CKM parameters this strategy is no longer 
necessary although it gives presently a very similar result \cite{Buras:2012ts}.

As far as the direct lattice determination of $f_{B_s}$ is concerned, an
impressive progress has been made in the last years 
\cite{Davies:2012qf,Dimopoulos:2011gx,McNeile:2011ng,Bazavov:2011aa,Blossier:2011dk,Na:2012kp,%
Gamiz:2009ku,Bouchard:2011xj}. These results are summarized in \cite{Davies:2012qf} and included 
in the world average $f_{\Bs} = (227.6 \pm 5.0)$ MeV \cite{Laiho:2009eu}. Using this result at 
face value we would get a total error on $\cBSM$ of $\pm 0.2 \times 10^{-9}$ in eq.~(\ref{eq:B0num_ref}).
However, given that this average is largely dominated by a single determination~\cite{McNeile:2011ng}, 
and given that all the other unquenched estimates of  $f_{B_s}$ have errors of about $\pm10$~GeV, 
we believe that a $\pm 8$~MeV error on $f_{\Bs}$ -- that we deduce from the spread of the central 
values -- is a more conservative estimate of the present uncertainty.

\section{Soft-photon corrections and the experimental branching ratio}\label{sec:IRcorr}

\subsection{Soft-photon corrections} \label{sec:omega}

As anticipated in the introduction, switching on electromagnetic interactions the $B_s\to \mu^+\mu^-$
transition is unavoidably accompanied by real photon emission. On general grounds we can distinguish
two types of radiation: bremsstrahlung and direct emission. The former is largely dominant for sufficiently 
small photon energies, can be summed to all orders in the soft-photon approximation, and leads to a 
multiplicative correction factor with respect to the non-radiative rate. On the contrary, the 
direct-emission component vanishes in the limit of small photon energies and represents a background 
for the extraction of short-distance information on the $B_s\to \mu^+\mu^-$ amplitude. A tight cut on the 
$\mu^+\mu^-$ invariant mass ($m_{\mu^+ \mu^-}$) close to $m_{B_s}$ allows us to treat radiative corrections 
in the soft-photon approximation and to suppress the background due to the direct-emission component.

In the soft-photon approximation ($\Emax \ll m_{B_s}/2$), the correction factor defined in 
eq.~(\ref{eq:defomega}), relating the photon-inclusive rate to the theoretical non-radiative rate, can 
be expressed as~\cite{YFS,WeinbergIR,IsidoriIR}
\bea
\label{eq:omega}
\om(\Emax) &=&  \om_{\rm IB}(\Emax) \times 
\left[ 1 + O\left(\frac{\alpha_{\rm em} }{\pi} \right) \right]~, \\
\label{eq:omegaIB}
\om_{\rm IB}(\Emax) &=& \left( \frac{2 \Emax}{m_{B_s}} \right)^{\frac{2\alpha_{\rm em} }{\pi} b}~,
\eea
where $\alpha_{\rm em}=1/137.036$ is the fine-structure constant and
\beq
b \equiv - \left[ 1 - \frac{1}{2 \beta_{\mm}} \ln \left( \frac{1+\beta_{\mm}}{1-\beta_{\mm}}\right) \right]
~, \qquad \beta_{\mm} = \left[ 1 - \frac{4 m_{\mu}^4}{(m^2_{\mu^+ \mu^-} - 2 m_{\mu}^2)^2}\right]^{1/2}~.
\eeq
The term  $\om_{\rm IB}(\Emax)$   takes into account the emission of an arbitrary number of real photons,
with maximal energy in the meson rest frame less or equal to 
\be
\Emax  =  \frac{m_{B_s}^2 - m^2_{\mu^+ \mu^-}}{2 m_{B_s}}~,
\ee
together with the corresponding virtual corrections: infrared divergences of real and virtual contributions 
cancel out leading to this universal correction factor.
The $O(\alpha_{\rm em}/\pi)$ term in eq.~(\ref{eq:omega}) represents the subleading  model-dependent 
contribution due to infrared-finite virtual corrections and due to the residual contribution of real 
emission, that vanishes in the limit  of vanishing photon energy. For $\Emax \approx 60$ MeV, the universal 
term yields
\beq
\om_{\rm IB}(\mbox{60 MeV}) \approx 0.89~,
\eeq amounting to a $\approx  11\%$ suppression of the non-radiative rate, whereas the  $O(\alpha_{\rm em}/
\pi)$ term is expected to be below the $1\%$ level, as discussed below. 

The normalization of $\Emax$ in  $\om_{\rm IB}(\Emax)$ is, in principle, arbitrary: different values lead 
to a redefinition of the  $O(\alpha_{\rm em}/\pi)$ finite term in eq.~(\ref{eq:omega}). Following 
Ref.~\cite{IsidoriIR}, we normalize $\Emax$ to its kinematical limit ($m_{B_s}/2$) in order to minimize 
the residual finite corrections.  The latter can be decomposed into the following three parts.

\begin{enumerate}
\item[I.] The residual real and virtual corrections in the absence of direct couplings of the meson to the 
photon. With the normalization adopted for $\om_{\rm IB} (\Emax)$, these corrections amount to 
$5\alpha_{\rm em}/(4\pi) \approx 0.3\%$, corresponding to the electromagnetic corrections for the decay 
of a point-like meson fully inclusive of bremsstrahlung radiation~\cite{Janot:1989jf}.

\item[II.] The virtual structure-dependent terms (due to effective non-minimal couplings of the meson to 
the photon). These terms are model dependent; however, they must respect the helicity suppression of the 
non-radiative amplitude and do not contain large logs. As a result, they are expected to be of the 
same size as those in point I.

\item[III.] The real contribution of the direct-emission amplitude. Since the direct-emission amplitude 
for $B_s \to \mu^+\mu^-\gamma$ is not helicity suppressed, it may represent a significant contribution 
if the $\Emax$ cut is not tight enough. However, according to Low's theorem~\cite{Low:1958sn}, the 
interference of bremsstrahlung and direct-emission amplitudes leads to a correction to the rate that
vanishes at least quadratically with the photon energy cut. From a naive dimensional analysis, the 
relative direct-emission contamination, for a given $\Emax$ cut, is
\be
\delta_{\rm DE} \leq 2 b \left( \frac{2 \Emax}{m_{B_s}} \right)^2 \times \left[ \frac{\alpha}{\pi} 
\frac{\cB(B_s \to   \mu^+\mu^-\gamma)_{\rm DE}}{\cBSM } \right]^{1/2}~,
\ee
where $\cB(B_s \to \mu^+\mu^-\gamma)_{\rm DE}$ represents the genuine direct-emission branching fraction.
According to the estimates in the literature (see Ref.~\cite{Melikhov:2004mk} and references therein)
the latter is $O({\rm few}\times 10^{-8}$). Then, if we assume 
$\cB(B_s \to \mu^+\mu^-\gamma)_{\rm DE} < 10^{-7}$ as a conservative estimate, we find that this 
relative correction is below $1\%$  for $\Emax < 100$~MeV.
\end{enumerate}

\subsection{Connecting the experimental with the theoretical branching ratio} \label{sec:BR_exp_th}

In order to obtain a theoretical prediction for the decay rate accessible in experiments, the 
last point we need to take into account is the effect of the non-vanishing width difference 
$\Delta\Gamma_s$, that has been measured recently rather precisely \cite{LHCb-DGammas}.
Following Ref.~\cite{Fleischer1}, we assume that what is presently measured by the LHC experiments 
is the flavor-averaged time-integrated distribution,
\be
\langle \cB(B_s \to f ) \rangle_{[t]}  =  \frac{1}{2} \int_0^t dt^\prime  \left[ \Gamma(B_s (t^\prime)
\to f ) + \Gamma(\bar B_s (t^\prime)\to f ) \right],
\ee
where $\Gamma(B_s(t^\prime)\to f)$ denotes the decay distribution, as a function of the proper time 
($t^\prime$), of  a $B_s$ flavor eigenstate at initial time  (and correspondingly for $\bar B_s$). 
Furthermore one defines
\be
\Gamma_s = \frac{1}{\tau_{B_s}} = \frac{1}{2} \left( \Gamma^H_s +\Gamma^L_s \right)~, 
\qquad  y_s = \frac{ \Gamma^L_s - \Gamma^H_s }{ 2 \Gamma_s } = 0.088 \pm 0.014~,
\ee
with $\Gamma^{H,L}_s$ the total decay widths of the two mass eigenstates. As discussed in 
 Ref.~\cite{Fleischer1}, the time-integrated distribution is related to the flavor-averaged rate at 
$t=0$ by 
\be
\langle \cB(B_s \to f ) \rangle_{[t]}  =  \kappa^f(t, y_s) \langle \cB(B_s \to f ) \rangle_{[t=0]} 
\equiv \kappa^f(t,y_s)  \frac{\Gamma(B_s \to f ) + \Gamma(\bar B_s \to f ) }{2 \Gamma_s}~,
\ee
where $\kappa^f(t,y_s)$ is a model- and channel-dependent correction factor. 

For the $\mu^+\mu^-$ final state (inclusive of bremsstrahlung radiation) the SM expression of the 
$\kappa^f(t,y_s)$ factor is~\cite{Fleischer2}
\be
\kappa_{\rm SM}^{\mu\mu}(t,y_s) = \frac{1}{ 1-y_s}\left[ 1 - e^{-t/\tau_{B_s}} \sinh
\left(\frac{y_s t}{\tau_{B_s}} \right) - e^{-t/\tau_{B_s}} \cosh\left(\frac{y_s t}{\tau_{B_s}} 
\right) \right] ~ \stackrel{~t\ \gg\ \tau_{B_s}~}{\longrightarrow}  ~ \frac{1}{ 1-y_s}~,
\ee
while the flavor-averaged branching ratio at $t=0$ is the quantity evaluated in the previous two 
sections. Putting all the ingredients together we then arrive at the following expression
\be
\langle \cB(B_s \to \mu^+\mu^-(\gamma) ) \rangle^{\rm SM}_{[t, \Emax]} = 
\kappa_{\rm SM}^{\mu\mu}(t,y_s) \times \om(\Emax) \times \cBSM~,
\label{eq:final}
\ee
for the quantity accessible in experiments. 

A few comments are in order:

\begin{itemize}

\item{}
The quantity which is more interesting for precise SM tests, and which can easily be affected by 
new-physics contributions, is $\cB^{(0)}$. The correction term $\om(\Emax)$ is insensitive to new 
physics, while $\kappa^{\mu\mu}(t,y_s)$ can deviate from its SM expression only in the presence 
of new-physics models with new CP-violating phases and/or non-standard short-distance operators 
contributing to the $B_s\to\mu^+\mu^-$ amplitude~\cite{Fleischer2}. 
Most importantly, the two correction terms $\om(\Emax)$ and  $\kappa^{\mu\mu}(t,y_s)$  need to be 
convoluted with the experimental efficiencies on $\Emax$ and $t$, and,  in principle, can even be 
determined experimentally up to their overall normalization (although an experimental determination 
of both these terms will become feasible only with a significant sample of $\Bsmumu$ events).
As a result, we encourage the experimental collaborations to directly provide a 
determination of $\cB^{(0)}$, already corrected for these two terms.
\item{}
Since $\om_{\rm IB}(m_{B_s}/2)=1$ and $\kappa_{\rm SM}^{\mu\mu}(t,y_s) \approx t/\tau_{Bs}$ for 
$t \ll \tau_{Bs}$, the theoretical quantity $\cBSM$ can be identified with the SM branching ratio 
of a flavor-tagged $B_s$ state at small times, fully inclusive of bremsstrahlung radiation only.
We stress once more that the necessity to include the correction factor $\om(\Emax)$ does depend 
on the treatment of the electromagnetic radiation in the measurement. For instance, in the recent 
LHCb result~\cite{Aaij:2012ac}, the signal is simulated fully inclusive of bremsstrahlung radiation 
and the correction term $\om(\Emax)$ (properly convoluted) is taken into account in the signal 
efficiency.\footnote{~We thank Tim Gershon and Matteo Palutan for useful discussions
regarding this point.}
\item{}
Finally, it is interesting to note that for the experimental choice of $\Emax$  applied by LHCb
($\Emax \approx 60$~MeV)~\cite{Aaij:2012ac}, and for $t\gg \tau_{Bs}$, the two correction terms in 
eq.~(\ref{eq:final}) tend to compensate each other to a large extent.

\end{itemize}

\section{\boldmath The $B_d\to \mu^+\mu^-$ decay} \label{sec:Bdmumu}

The corresponding analysis of the $B_d\to\mu^+\mu^-$ decay is a straightforward 
generalization of the one just presented for $B_s\to\mu^+\mu^-$. As far as the 
three items listed in the Introduction are concerned, the following 
comments suffice:
\begin{itemize}
\item
Our analysis of short-distance NLO QCD and NLO electroweak corrections 
{(sec. \ref{sec:nonrad})} remains unchanged. In particular our conjecture, 
summarized by eqs. (\ref{CONJ1}) and (\ref{eq:Y0prescription}), {applies identically 
in this case}. {What is trivially modified in the basic 
expression in eq. (\ref{eq:BR0}) are the initial-state constants $m_{\Bs}$, $\tau_{\Bs}$,
$f_{\Bs}$ and the CKM coupling, as now the index $s$ is replaced by $d$.}
These new input parameters are given in Table~\ref{tab:input}.
\item
The soft-photon corrections {(sec. \ref{sec:IRcorr}) remains likewise unchanged, 
as the $B_s$ and $B_d$ masses are very close to each other.}
\item
{The effect of $\Delta\Gamma_d$ (see sec. \ref{sec:BR_exp_th}) is negligible.}
\end{itemize}
Thus the final expression in eq. (\ref{eq:final}) is replaced by
\be
\langle \cB(B_d \to \mu^+\mu^-(\gamma) ) \rangle^{\rm SM}_{[t, \Emax]} = 
\om(\Emax) \times \cBSMd~.
\label{eq:finald}
\ee

Furthermore, using the input in  Table~\ref{tab:input} we find {as the analogues of 
eqs.} (\ref{eq:B0num_ref}) and (\ref{eq:BR0par}) the following results
{
\be
\label{eq:B0num_refd}
 \cBSMd =(1.07 \pm 0.10)  \times 10^{-10}~,
\ee
\bea
\label{eq:BR0pard}
&& \cBSMd =  {1.0659} \times 10^{-10} \times \left( \frac{M_t}{173.2~\gev} \right)^{3.07}
\left( \frac{f_{B_d}}{190~\mev} \right)^2
\left( \frac{\tau_{B_d}}{1.519~\rm{ps}}\right)
\left| \frac{V^*_{tb} V^{\phantom{*}}_{td} }{8.7 \times 10^{-3}} \right|^2 \nonumber  \\
&& \qquad =   \left(1.07 \pm 0.05 \pm 0.09_{f_{B_d}}\right)   \times 10^{-10}~.
\eea
The result in eq. (\ref{eq:B0num_refd}) is our reference value for the non-radiative branching 
ratio of the $B_d\to\mu^+\mu^-$ decay. In addition, similarly as for the $B_s$ case, 
eq. (\ref{eq:BR0pard}) illustrates the impact of the various inputs on the quoted central 
value and error. In the second line of this equation we have explicitly separated out the 
contribution to the error due to $f_{B_d}$, which is the most relevant source of uncertainty.}

{
Finally, eqs. (\ref{eq:BR0par}) and (\ref{eq:BR0pard}) translate straightforwardly into
a prediction for the ratio of the non-radiative branching ratios, $\BRBsmumu / \BRBdmumu$.
Using $f_{\Bs}/f_{\Bd} = 1.195$ from the separate constants in our table, and indicating
its {\em relative} error as $\sigma^{r}_{f_{s/d}}$, one easily finds
\be
\frac{\cBSM}{\cBSMd} ~=~ 30.35 \left( 1 \pm 0.06 \pm 2 \sigma^r_{f_{s/d}} \right) ~.
\ee
}

\section{Summary}

In the present paper we have presented a comprehensive discussion of all the effects that 
are expected to have a significant impact on the SM prediction of $\cB(B_s \to \mu^+ \mu^-)$.
By this we mean that we expect residual uncertainties to be negligible with respect to the 
foreseeable experimental accuracy. 
In particular we have discussed the effects of NLO electroweak corrections, and 
of the correspondence between the theoretical branching ratio and the experimental one, 
focusing on the effect of soft bremsstrahlung photons. Our main messages from this analysis 
are as follows:
\begin{itemize}
\item
The main uncertainty in the prediction of the non-radiative branching ratio $\cBSM$, by 
definition independent of soft-photon corrections, still originates in $f_{B_s}$, that enters 
$\cBSM$ quadratically. However, the impressive progress made by lattice QCD evaluations 
in the last two years makes this error as low as O(5\%) at the level of the branching ratio.
\item
At this level of accuracy it becomes essential to perform a complete calculation of NLO 
electroweak corrections to $\cBSM$, that in this case are at present known only in the 
large-$m_t$ limit \cite{BB-large-mt}. An explicit evaluation of these corrections is the only 
means by which the renormalization-scheme dependence due to the scheme choice for electroweak 
parameters like $\sin^2 \theta_W$, can be reduced to a really negligible level. Using the 
large-$m_t$ limit approximation we estimate the present error due to unknown NLO electroweak 
corrections to be $\pm 3\%$. The recently performed complete NLO analysis of these corrections 
in the case of $K\to\pi\nu\bar\nu$ decays reduced the corresponding uncertainty down to 
per mil level \cite{Brod:2010hi}.
\item 
Anticipating the structure of complete NLO electroweak corrections in $B_s\to \mu^+\mu^-$ 
to be similar to the known case of $K\to\pi\nu\bar\nu$, we have conjectured that the most 
reliable value for $\cBSM$ can be obtained by choosing $\sin^2\theta_W$ in the $\MSbar$ scheme, 
the top mass in the $\MSbar$ scheme only as far as QCD corrections are
concerned, and taking the short-distance function {to be the sum of the LO one 
and of QCD corrections only.} This simple prescription leads to a prediction for the $\cBSM$
\be\label{eq:B0SM}
{
\cBSM \equiv \cB(B_s \to \mu^+ \mu^-)^{\rm SM} = (3.23 \pm 0.27) \times 10^{-9}.
}
\ee
This is our reference value for the non-radiative branching ratio. 
This result 
is lower by $2\%$ to $7\%$, depending on the scheme considered, with respect to 
the estimates including NLO two-loop electroweak corrections in the large $m_t$-limit, 
implying an anticipated significant role of missing NLO electroweak corrections.
Nonetheless, we have argued that, within our prescription, electroweak corrections are 
plausibly tiny.
\item
In connection with this prediction, formula (\ref{eq:BR0par}) should allow to monitor 
how the central value for $\cB(B_s\to\mu^+\mu^-)$ changes as a function of its main
parametric dependencies.
\item
We have addressed the issue of the correspondence between the initial and the final state 
detected by the experiments, and those used in the theoretical prediction. In particular,
we have focused on the effect of the soft radiation, that has not been discussed for the
$\Bsmumu$ mode in the previous literature, and that can lead to O(10\%) corrections to 
the decay rate. We have argued that, if the sum of the energies of the undetected photons
is small enough, the dominant effect is due to the correction factor in eq. (\ref{eq:omegaIB}),
and we have discussed the expected magnitude of residual effects. This correction may provide
a useful comparison yardstick against a more accurate Monte Carlo estimate, where non-uniform
experimental efficiencies are properly taken into account.
\item
Including the effect of $\bar B_s - B_s$ oscillation, recently pointed out in 
\cite{Fleischer1,Fleischer2} and also of O(10\%), we arrive finally 
at the relation (\ref{eq:final}), 
that connects the theoretical with the experimental branching ratio.
\item
{
A completely analogous procedure, applied to the $B_d \to \mu^+\mu^-$ decay, leads to
\be\label{eq:B0SMd}
{
\cBSMd \equiv \cB(B_d \to \mu^+ \mu^-)^{\rm SM} = (1.07 \pm 0.10) \times 10^{-10},
}
\ee
allowing, along with formula (\ref{eq:BR0pard}), to monitor how the central value for 
$\cB(B_d\to\mu^+\mu^-)$ changes as a function of its main parametric dependencies.}
\end{itemize}

We are looking forward to a precise measurements of $\cB(B_s\to\mu^+\mu^-)$ 
and  $\cB(B_s\to\mu^+\mu^-)$, with the hope
that they  will disagree with $\cBSM$ and $\cBSMd$ in eqs. (\ref{eq:B0SM}) and 
(\ref{eq:B0SMd}), respectively.

\section*{Acknowledgements}
We thank Damir Becirevic, Joachim Brod, Fulvia de Fazio, Tim Gershon, Martin Gorbahn, 
Mikolaj Misiak, Matteo Palutan, Alexey Petrov and Emmanuel Stamou for useful comments and discussions. 
This work was supported by the EU ERC Advanced Grant FLAVOUR (267104), and by MIUR under 
contract 2008\-XM9HLM.

{\footnotesize
\bibliographystyle{My}
\bibliography{BGGI}
}

\end{document}